\documentclass{ws-procs9x6}

\begin{document}

\title{Detector performance of the NEWAGE experiment}

\author{
Kentaro~Miuchi$^a$,
Kaori~Hattori$^a$, 
Shigeto~Kabuki$^a$, 
Hidetoshi~Kubo$^a$,
Shunsuke~Kurosawa$^a$, 
Hironobu~Nishimura$^a$, 
Yoko~Okada$^a$, 
Atsushi~Takada$^a$,
Toru~Tanimori$^a$,
Ken'ichi~Tsuchiya$^a$,
Kazuki~Ueno$^a$,
Hiroyuki~Sekiya$^b$, 
Atsushi~Takeda$^b$}

\address{
Cosmic-Ray Group, Department of Physics, Graduate School of Science, 
Kyoto University Kitashirakawa-oiwakecho, Sakyo-ku, Kyoto, 606-8502, Japan$^a$ 
E-mail: miuchi@cr.scphys.kyoto-u.ac.jp}

\address{
 Kamioka Observatory, ICRR, The Univ. of Tokyo
Higashi-Mozumi, Kamioka cho, Hida 506-1205 Japan $^{b}$}

\begin{abstract}
\end{abstract}
NEWAGE(NEw generation WIMP search with an Advanced Gaseous tracking device Experiment) project is a direction-sensitive dark matter search experiment with a gaseous micro time-projection-chamber($\mu$-TPC).
We report on the performance of the $\mu$-TPC with a detection volume of 
23 $\times$ 28  $\times$ 30 $\rm cm^3$ operated with a  carbon-tetrafluoride ($\rm CF_4$) of 0.2 bar.
\keywords{
Time projection chamber; Micro-pattern detector; dark matter}

\bodymatter

\section{Introduction}\label{section:Introduction}

Weakly interacting massive particle(WIMPs) are thought to be one of the 
most plausible candidates of the dark matter. 
Most of the dark matter search experiments are designed to measure 
only the energy deposition 
on the nucleus 
by a WIMP-nucleus scatterings.
Because the amplitude of an annual modulation signal is 
only a few $\%$ in the rate, 
positive signatures of the WIMPs are 
very difficult for detection with only the energy information.
Owing to the motion of the solar system with respect to the galactic halo,
the direction-distribution of the WIMP velocity observed 
at the earth is expected to show an asymmetry like a wind of WIMPs. 
Attempts to detect a positive signature of WIMPs
by measuring the recoil angles have been carried 
out \cite{ref:DAMA_aniso,ref:Buckland_PRL,ref:APP_CH4,ref:DRIFT2_NIM,ref:Sekiya_IDM2004} 
ever since it was indicated to be an alternative and a reliable 
method \cite{ref:Spergel_WIMP}.
Gaseous detectors are one of the most appropriate 
devices for detecting this WIMP-wind.
DRIFT project has performed underground runs for more than two years
with a 1m$^3$ time projection chamber (TPC)
filled with a low pressure $\rm CS_2$ gas\cite{ref:DRIFT2_NIM}.
We proposed to use a carbon-tetrafluoride ($\rm CF_4$) as a 
chamber gas of our time projection chambers with a micro-pixel chamber 
readout ($\mu$-TPC) 
aiming to detect WIMPs via spin-dependent (SD) interactions\cite{ref:Miuchi_PLB}. In this paper, the performance of the $\mu$-TPC is described.

\section{Measurements}
\subsection{Micro-TPC}
\label{section:micro-TPC}
A $\mu$-TPC is a time projection chamber with a micro pixel chamber 
($\mu$-PIC\cite{ref:uPIC}) 
readout, developed for the detection of tracks 
of charged particles with fine spatial resolutions
\cite{ref:Sekiya_PSD}. 
A $\mu$-PIC is a gaseous two-dimensional position-sensitive detector 
manufactured by the printed circuit board (PCB) technology. 
With the PCB technology, large-area detectors can, in principle, be 
mass-produced, 
which is an inevitable feature for a dark matter detector.
The pixel-pitch of the $\mu$-PIC is 400 $\mu$m and the detection area is 
31$\times$31 cm$^2$.
We had studied the performance of a small size
(10 $\times$ 10 $\times$ 10 $\rm cm^3$ ) $\mu$-TPC
with a 0.2 bar $\rm CF_4$ gas\cite{ref:Sekiya_PSD}.
We then developed a large-volume $\mu$-TPC with a detection volume of 
23 $\times$ 28  $\times$ 30 $\rm cm^3$, and 
studied its fundamental properties 
with an Ar-$\rm C_2 H_6$ gas mixture 
at a normal pressure\cite{ref:Miuchi_IWORID8}.
The data acquisition system is described in Ref. \cite{ref:Kubo_IEEE}.
We will describe the performance of the 
$\mu$-TPC with a 0.2 bar $\rm CF_4$ gas in the following subsections.


\subsection{Energy calibration}
\begin{figure}[h]
  \begin{center}
    \psfig{file=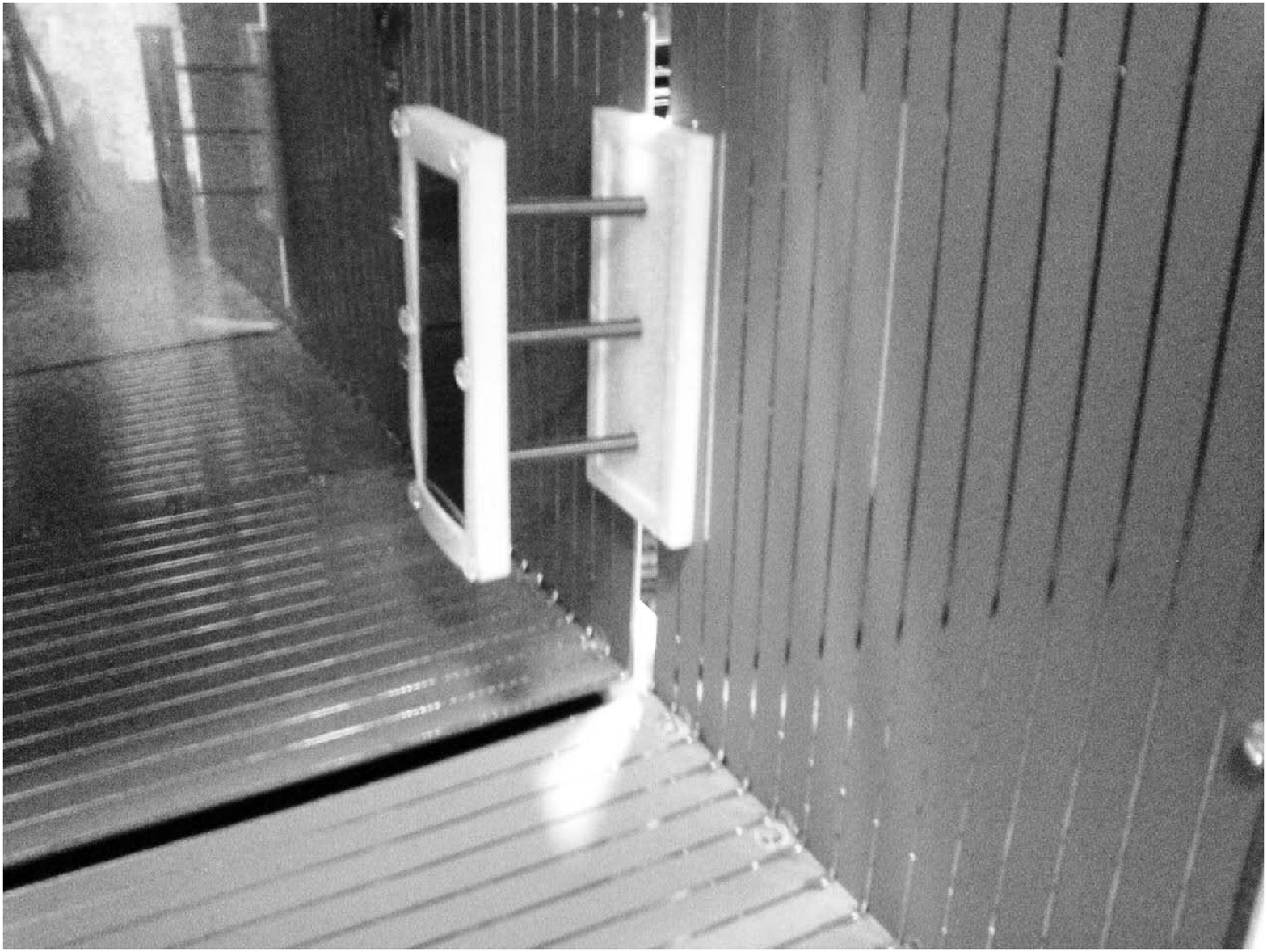,width=2in}
    \psfig{file=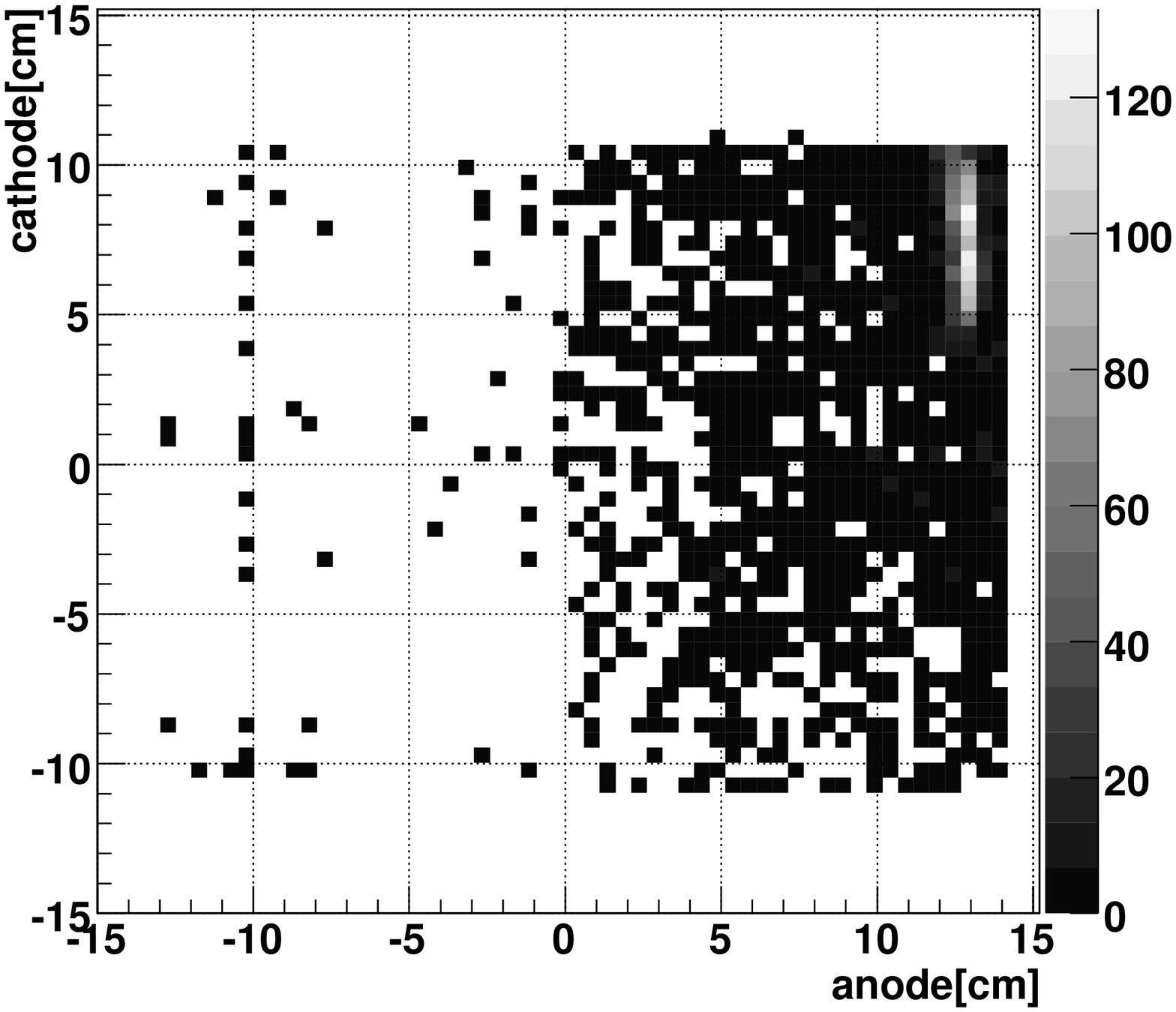,width=2in}
    \caption{Picture of the glass plate coated with a boron layer(right) and 
an integrated event display of a calibration measurement.}
  \label{fig:B10image}
  \end{center}
\end{figure}

We calibrated the energy of the $\mu$-TPC with $\alpha$ particles
generated by the $\rm ^{10}B (n,\alpha)^7Li$(Q=2.7MeV) reaction.
We set a glass plate with a size of $\rm 27\times 70\times 1mm^3$ 
coated with a thin 0.6 $\mu$m $\rm ^{10}B$ layer 
in the $\mu$-TPC.
The picture of the boron-coated glass set in the $\mu$-TPC is shown in 
the left panel of Fig. \ref{fig:B10image}.
Fast neutrons from $\rm ^{252}Cf$ 
were moderated and the thermalized neutrons were captured by the 
$^{10}$B layer.
Alpha particles are emitted from the layer and the 
integrated two-dimensional image
is shown in the right panel of Fig.~\ref{fig:B10image}.
It is seen that 
the $\alpha$ particles from the $^{10}B layer$ are detected 
and the corresponding position has a high counting rate.  

\begin{figure}[h]
   \begin{center}
\psfig{file=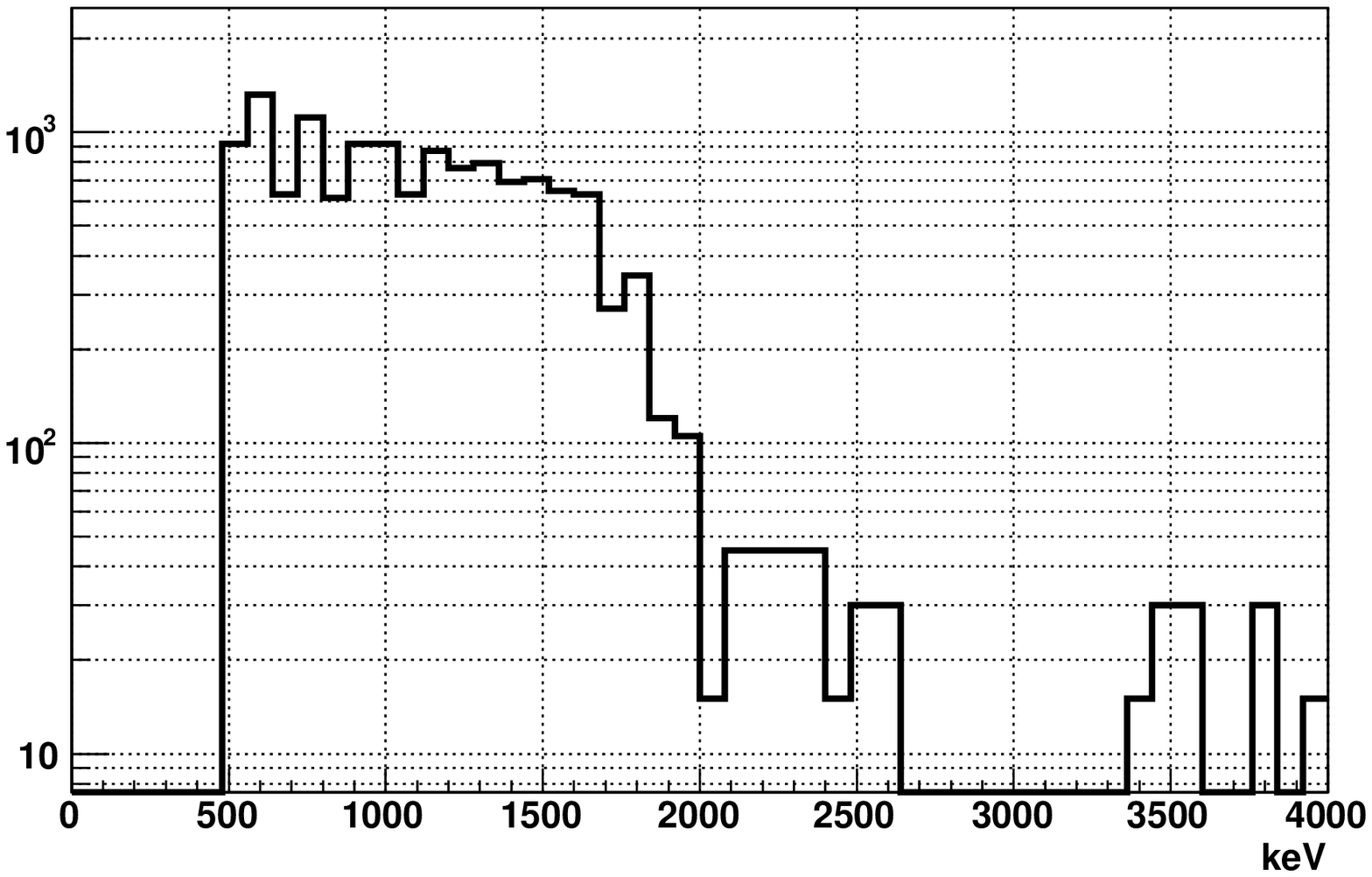,width=2in}
\psfig{file=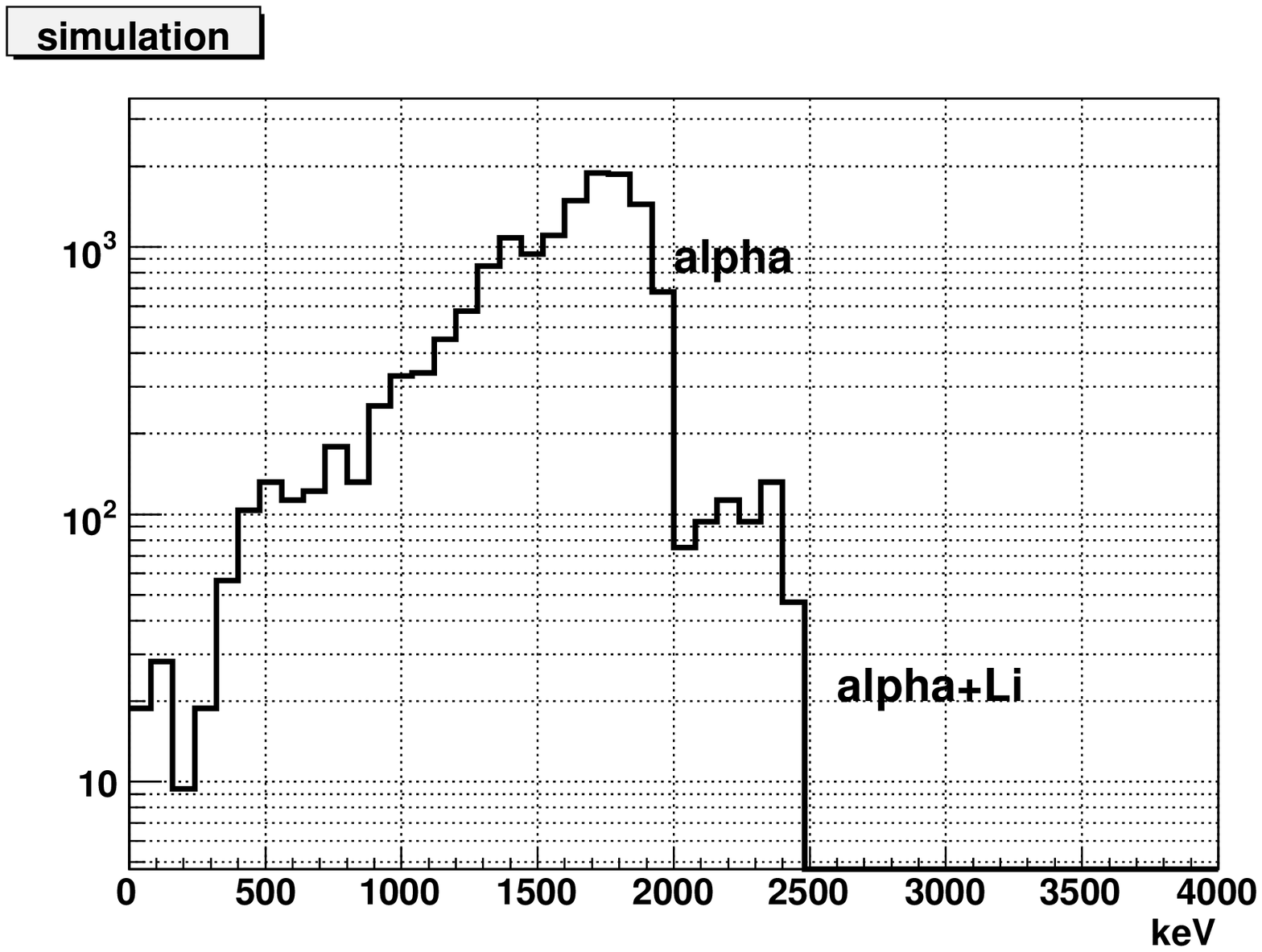,width=2in}
\caption{Measured(left) and simulated spectra of the calibration   }
 \label{fig:B10spec}
  \end{center}
   \end{figure}

The measured and simulated spectra are shown in Fig \ref{fig:B10spec}.
The edge which corresponds to the full energy deposition of the 
$\alpha$ particle and the $\alpha$+Li are seen in the both spectra.
The elastic scatterings of the fast neutrons from the $^{252}$Cf source 
gives an increase of the lower energy part of the measured energy spectrum.

Alpha peaks(5.6 MeV, 6.1 MeV, 7.2 MeV) 
form the decays of the radon daughters are also used 
for the energy calibration.
The energy resolution was also measured with these alpha peaks 
and the energy resolution was 50$\%$(FWHM) in the high energy(3-8 MeV) region.

\subsection{Absolute detection efficiency of nuclear recoils}
The detection and the event-selection efficiency was measured by 
irradiating the fast neutrons from $^{252}$Cf. 
Here the events which satisfied the following three conditions 
were selected as a nuclear recoil events.
\begin{itemize}
\item   The track is in the fiducial volume (21.5 $\times$ 22 $\times$ 31 $\rm cm^3$).
\item   The track has more than three digital hits.
\item   The track is shorter than 1cm. 
\end{itemize}

\begin{figure}[h]
   \begin{center}
     \psfig{file=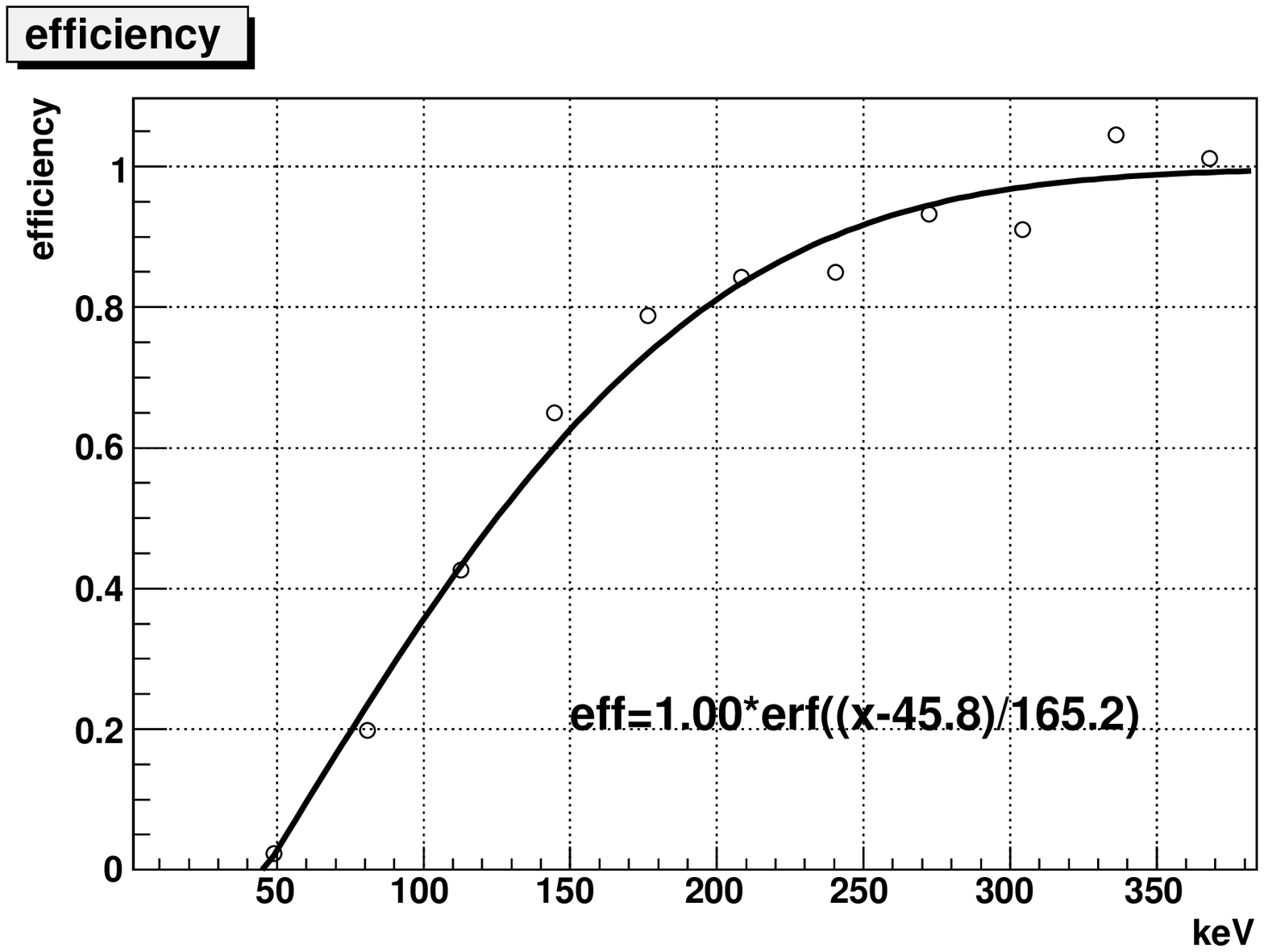,width=2in}
     \psfig{file=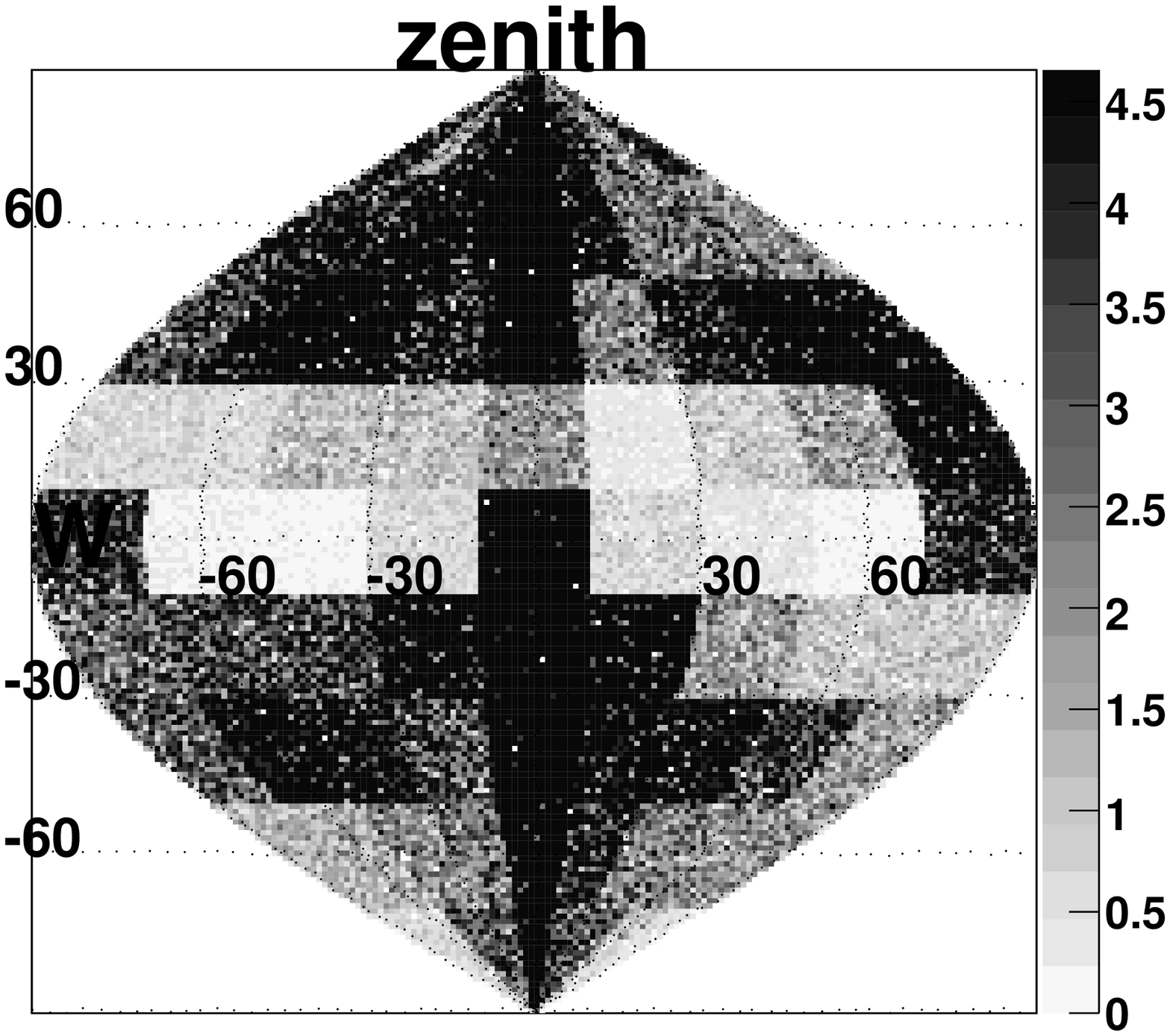,width=2in}
     \caption{Absolute detection efficiency of the nuclear recoils (left) and the relative direction-dependent response. }
     \label{fig:eff}
  \end{center}
   \end{figure}

The first requirement is the fiducial cut to reject the protons from the drift wall, the second is for the direction determination, and the third one is for 
the gamma-ray rejection. 
The detection and the selection efficiency was about 40$\%$ at 100 keV and the efficiency in the energy region of 100-400keV (DM energy region) $\epsilon$ was 
fitted by $\epsilon=1.0\cdot erf((E-45.8)/165.2)$,
where $erf(x)=\frac{2}{\sqrt{\pi}}\int^{x}_{0}exp(-x^2)dx$ is the error function and $E$ is the detected energy.
The measured efficiency and the best fit function are shown in the left panel of Fig. \ref{fig:eff}

\subsection{Direction-dependent detector response}
The direction-dependent detector response is one of the most important
properties for the direction-sensitive dark matter search.
We irradiated the fast neutrons from various position to generate 
''a uniform recoils''.
We determined the direction of the selected events 
by fittings the digital hits
and the direction-map is shown in the right panel of Fig. \ref{fig:eff}.
Because we don't detect the head-and-tail of the tracks, 
this mas is restricted to the half sky, {\it i. e.} -90$^{\circ}<$ azimuth 
$<$90$^{\circ}$ and -90$^{\circ}<$ zenith $<$90$^{\circ}$
The response is not very flat nor symmetry because of the 
inhomogeneity of the detector.
A much more flat response is expected for the next detector with an 
improved manufacture technologies.

\subsection{Gamma-ray rejection}
\begin{figure}[h]
   \begin{center}
     \psfig{file=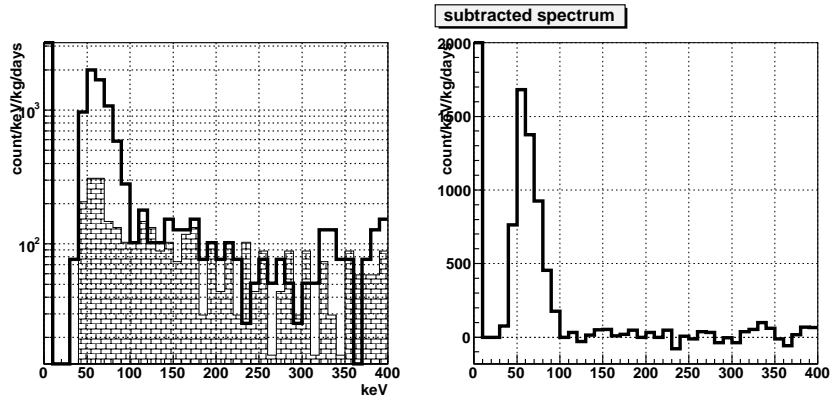,width=4.5in}
\caption{
The energy spectra measured with(solid) and without(hatched) the gamma-ray source(left). The background-subtracted(=solid-hatched) spectrum(right).   }
 \label{fig:gammarej}
  \end{center}
   \end{figure}

It is one of the outstanding advantage of the gaseous TPC that
the track-length and the energy-deposition 
correlation provides a strong gamma-ray rejection\cite{ref:Sekiya_PSD}.
We measured the gamma-ray rejection factor by irradiating the 
gamma-rays from a $\rm^{137}Cs$ source.
Measured spectra with and without the gamma-ray source and the background-subtracted spectrum are show in Fig.~\ref{fig:gammarej}
We define the gamma-ray miss-identification factor 
(or the electron-detection efficiency) 
$f_{e}$ as
\begin{equation}
f_{e}\equiv \frac{R_{\gamma}-R_{\rm BG}}{R_{\rm sim}}, 
\end{equation}
where 
$R_{\gamma}$ and $R_{\rm BG}$ are the
the measured counting rate DM energy region 
with and without the gamma-ray source 
and $R_{\rm sim}$, is the expected rate of the 
electrons which has a initial energy in the DM energy region.
The result was $f_e< 2.0 \times 10 ^{-4}$.
This ''rejection factor''  was 
restricted by the background neutrons in the surface laboratory.


\section{Summary}
We measured the performance of the $\mu$-TPC and 
obtained the following results.
\begin{itemize}
\item The detector is calibrated with $\alpha$-particles.
\item The absolute efficiency is about 0.4$\%$ at 100keV.
\item The electron miss-identification probability is less than $2\times10^{-4}$ in the 100-400 keV energy range. 
\item The direction dependent response is measured.
After a pilot run in the surface laboratory, we will start an 
underground measurement in Kamioka Observatory in January, 2007. 
\end{itemize}

\section*{Acknowledgments}
This work was supported by a Grant-in-Aid in Scientific Research of 
the Japan Ministry of Education, Culture, Science, Sports, Technology; 
research-aid program of Toray Science Foundation;
and Grant-in-Aid for the 21st Century COE 
''Center for Diversity and Universality in Physics''.

\end{document}